\documentclass[12pt]{article}

\usepackage{amsthm}
\usepackage{amsmath}
\usepackage{amsfonts}
\usepackage{amssymb}
\usepackage{amscd}
\usepackage{mathrsfs}
\usepackage[all]{xypic}
\usepackage{url}
\usepackage{epsfig}

\newtheorem{thm}{Theorem}[section]
\newtheorem{cor}[thm]{Corollary}
\newtheorem{lem}[thm]{Lemma}
\newtheorem{prop}[thm]{Proposition}

\newtheorem{remark}[thm]{Remark}
\newtheorem{algorithm}[thm]{Algorithm}

\theoremstyle{definition}
\newtheorem{defn}[thm]{Definition}%[section]

\theoremstyle{definition}
\newtheorem{example}[thm]{Example}

\newcommand{\im}{\operatorname{im}}

\newcommand{\N}{\mathbb{N}}

\newcommand{\R}{\mathbb{R}}

\begin{document}
\title{The Cyclohedron Test for Finding Periodic Genes \\
 in        Time Course Expression Studies }
\author{Jason Morton, Lior Pachter, Anne Shiu and Bernd Sturmfels}
\maketitle
\begin{abstract}
The problem of finding periodically expressed genes from time course
 microarray experiments is at the center of numerous efforts to identify the molecular
 components of biological clocks. We present a new approach to this
 problem based on the cyclohedron test, which is a rank test inspired by recent advances in algebraic
 combinatorics. The test
  has the advantage of being robust to measurement errors, and
 can be used to ascertain the significance of top-ranked genes.
 We apply the test to recently published
 measurements of gene expression during mouse somitogenesis and find
 32 genes that collectively are
 significant. Among these are previously identified periodic genes
 involved in the Notch/FGF and Wnt signaling pathways, as well as novel
 candidate genes that may play a role in regulating the segmentation clock.
 These results confirm that there are an abundance
 of exceptionally periodic genes expressed during somitogenesis. The emphasis of this paper is on the 
 statistics and combinatorics that underlie the cyclohedron test and its implementation within a multiple testing framework.
\end{abstract}

\newpage

\section{Introduction}

The search for the molecular components of biological clocks is an
important first step towards understanding the regulatory mechanisms
underlying periodic behavior at the molecular level. Examples of clocks
that have been studied include the circadian clock \cite{McDonald}, the
respiratory cycle clock in yeast \cite{Klevecz, Spellman} and the segmentation
clock in vertebrates \cite{Pourquie}. In order to find clock-related
genes in a high-throughput fashion, time course array experiments are
performed to measure the expression levels of genes on a genome-wide
scale. This is followed by a statistical analysis to find periodically
expressed genes. The analysis is non-trivial for reasons that include
noisy measurements, variable times between experiments,
vague notions of periodicity, and loss of power due to multiple testing.

The question of how best to analyze cyclic time series is a
topic of extensive research in statistics \cite{Chatfield}.
Recent approaches, proposed in the context of microarray analysis include
splines and other curve approximations \cite{Luan,Storey},
methods based on signal processing techniques such as the Lomb-Scargle test \cite{LS}, and
non-parametric rank tests  \cite{Fink}.
 All of these methods address, to varying degrees, the
difficulties outlined above, and are sometimes developed in
response to specific needs dictated by individual experiments.

In this paper we introduce a new test for finding periodic genes.
Our method belongs to the family of convex rank tests
in \cite[Section 5]{GRT}. These tests were inspired by
{\em up-down analysis}, the method of \cite{Fink}.
They are based on recent advances in algebraic combinatorics,
namely the theory of {\em graph associahedra} \cite{Fomin, Hohlweg, Markl}.  The connection between rank tests and polytopes was first suggested in \cite{Cook}.
When using rank tests, an expression time-course is
represented by a permutation. This has the advantage of providing
robustness to noise, monotonic transformations, and uncertainty with respect to the underlying probability distributions, and the disadvantage of precluding a parametric analysis of the untransformed time courses.
In  up-down analysis, each permutation of $\{1,2,\ldots,n\} $ is
mapped to a sign vector, or {\em signature}, that records, for each adjacent pair on the $n$-path,
which of the two measurements is higher. Significance is determined by
counting the number of permutations that have an observed signature.

Our  {\em cyclohedron test} is based on a similar permutation count to that of
up-down analysis,
but the data points are now compared at longer range along the edges of the $n$-cycle.
The cyclohedron $C_n$ is the graph associahedron 
when the graph $G$ is the $n$-cycle, and 
the cyclohedron test is the greedy method for
linear programming on $C_n$. % visualized in Figure \ref{tubingsignature},
It is equivalent to the test denoted by
$\,\tau^*_{\mathcal{K}(G)} \,$ in \cite[Section 5]{GRT}.
Cyclohedra are also known as {\em Bott-Taubes polytopes},
and they play an important role in
representation theory \cite[Section 3.2]{Fomin},
combinatorics \cite{Hohlweg, Sandman}, and homotopy theory \cite{Markl}.
Connections to statistical learning theory were explored
and developed in \cite{semigraphoid, GRT}.

The cyclohedron test is explained in detail in Section \ref{sec:cyctest}.
Our presentation is elementary and self-contained.
In Section 3 we present a  method for assigning p-values to top-ranked groups of genes.
This is done within a multiple hypothesis testing framework,
which is compatible with any rank test
for permutation data, including up-down analysis. In Sections 5 and 6 we develop
the combinatorial details and efficient algorithms for the cyclohedron test.  Our {\tt R} code is available online,  and its use is described in the Appendix.

We apply the cyclohedron test to data reported in \cite{Science}, consisting of
17 distinct expression array experiments
from the presomitic mesoderm
tissue of mouse embryos.   These data were chosen because of the analyses already undertaken and the possibility for biological validation.
Results are discussed in Section 4. We find that
although the high-throughput array experiments are effective for finding groups
of genes likely to be involved with clock regulation, multiple testing
issues preclude the assignment of significance to any individual gene on the
basis of periodic-looking patterns alone.

\section{The cyclohedron test} \label{sec:cyctest}
The cyclohedron test is appropriate when seeking to determine whether a
 time course expression is periodic.  Within a single hypothesis
 setting, the null hypothesis states that a gene or other unit of
 interest does not exhibit cyclic expression.  The cyclohedron test
 provides a test statistic, which we call the {\em permutation count}, 
 that replaces this vague null hypothesis. 
The test applies to data vectors $v = (v_1,\ldots,v_n)$
whose coordinates are distinct real numbers.  The coordinates $v_i$ are
measurements of the same quantity at distinct points. In our applications, the 
ordering of each vector should be with respect to some `cyclic' time, so that
any  $v' = (v_i, v_{i+1}, \dots, v_n, v_1, \dots, v_{i-1})$ is an equally meaningful ordering.  For example, the data vectors $v$ we analyze in  Section \ref{Mouse} are ordered within a somite-formation cycle; so $v_j$ is a measurement taken before $v_{j+1}$ in the cycle, where $j+1$ is understood$\mod n$.

The following procedure computes, for any given data
vector $v$, its {\em signature} $\sigma(v)$
and its {\em permutation count} $\, \mathbf{c}(v) $. The signature
is an unordered
set $\sigma = \{\sigma_1,\sigma_2,\ldots,\sigma_{n-1}\}$ of
subsets of $ \{1,2,\ldots,n\}$ and the permutation count
is a positive integer.

 \bigskip
\begin{algorithm} {\rm (Cyclohedron test)}
\label{cyctestalg} \rm

{\em Input: } A vector $v = (v_1,\ldots,v_n)$ of distinct real numbers.

{\em Output:} The signature $\sigma = \{\sigma_1,\sigma_2,\ldots,\sigma_{n-1}\}$
 and the permutation count $\,\mathbf{c} \,$ for $\,v$.

\smallskip

Initialize $\mathbf{c} := 1$.

For $i$ from $1$ to $n{-}1$, do

\quad Initialize  $\sigma_i = \emptyset$, the empty set.

\quad Let $\delta_i$ be the unique index such that $v_{\delta_i}$ is the $i$-th
largest coordinate of $v$.

\quad Initialize $\,{\rm Left} := \emptyset \,$ and $\,{\rm Right} := \emptyset$.

\quad For $k$ from $1$ to $i {-} 1$, do

\qquad if $\sigma_k$ contains $\,  \delta_i {-} 1 \,$ (modulo $n$) then set $\,{\rm Left} := \sigma_k$,

\qquad if $\sigma_k$ contains $\,  \delta_i {+} 1 \,$ (modulo $n$) then set $\,{\rm Right} := \sigma_k$.

\quad
Set $\,\sigma_i \,\, := \,\, \{\delta_i\} \,\cup\,{\rm Right} \,\cup
{\rm  Left} \,\,$ and $\,\, \mathbf{c} \,:= \,\,\mathbf{c} \cdot \binom{|{\rm Right}| + {\rm |Left}|}{|{\rm Right}|}$.
\end{algorithm}

\smallskip

Let $\mathcal{C}_n$ denote the
set of all signatures $\sigma(v)$ as $v$ runs over $\R^n$.
Algorithm \ref{cyctestalg}
constructs not only the signature $\sigma$ and the permutation count
$\mathbf{c}$ but also the {\em descent order permutation} $\delta = (\delta_1,\delta_2, \ldots,\delta_n)$
of the data vector $v = (v_1,v_2,\ldots,v_n)$.
Since $\sigma(v)$ depends only on the descent order permutation $\delta$,
our algorithm specifies a map $\,\delta \mapsto \sigma\,$
from the symmetric group $\Sigma_n$ onto the set $\mathcal{C}_n$.
For $n \geq 4$, this map is not injective, and we are interested in the
cardinalities of the preimages.
For instance, the permutations $\delta = (1,3,2,4)$
and $\delta' = (3,1,2,4)$ have the same signature
$\,\sigma(\delta) = \sigma(\delta') = \bigl\{ \{1\}, \{3\}, \{1,2,3\} \bigr\}$.

The test statistic, the {\em permutation count} {\bf c}, is the number of permutations having the same 
signature as the permutation of interest. Significant data vectors have small test statistics, because it is unlikely that a random permutation will have a topographical map shared by few permutations.  The permutation count $\mathbf{c} = \mathbf{c}(v)$ has the following interpretation.
Suppose that an appropriate null data generating distribution for each data vector $v$ induces the uniform distribution on all descent order permutations $\delta$ in the symmetric group $\Sigma_n$. Note that this assumption is valid if the coordinates of the data vector are independent and identically distributed under the null distribution, 
so our test is therefore broadly applicable.  For each signature $\sigma \in \mathcal{C}_n$, let
$p(\sigma)$ denote the probability that
the signature $\sigma $ would be observed under such a null distribution.  The following proposition states that $p(\sigma)$ is the fraction of permutations
$\delta$ that map to~$\sigma$.

\begin{prop} \label{propscore}
The permutation count $\mathbf{c}$ computed by Algorithm \ref{cyctestalg}
depends only on the signature $\sigma$. It equals
the number of permutations $\delta$ that
are mapped to $\sigma$, and hence
$$\,\mathbf{c} \,\, = \,\, \mathbf{c}(\sigma) \quad  = \quad p(\sigma) \cdot n ! . $$
\end{prop}

\begin{proof}
For each $\sigma_i $  in the signature $\sigma$, at most two other
sets $\sigma_j$ and $\sigma_k$ are contained in
$\sigma_i$ and are maximal with this property.
Here $\sigma_j$ and $\sigma_k$ are
necessarily disjoint. The permutation count $\mathbf{c}$ is the product
of the corresponding binomial coefficients
$\,\binom{|\sigma_j \cup \sigma_k|}{|\sigma_j|}$.
It depends only on $\sigma$.
The second statement is proved by induction on $n$,
using the fact that any valid permutation
of $\sigma_j$ can be shuffled with any
valid permutation of $\sigma_k$, and augmented
by $\delta_i$, to get
a valid permutation for $\sigma_i$.
Carrying out this process until $i=n$, with  $\sigma_n = \{1,2,\ldots,n\}$,
yields precisely all permutations $\delta$ that have signature
 $\sigma$.  
\end{proof}

The other output of the algorithm, the signature $\sigma(v)$, can be viewed
as a topographic map on the $n$-cycle that captures the shape of the  data $v$. Algorithm \ref{cyctestalg}
is an iterative procedure for drawing this topographic map.
Namely, we encircle the vertices of  the $n$-cycle in decreasing order of their corresponding data vector coordinates, that is,
in the order $\delta_1, \delta_2, \dots, \delta_{n-1}$.  (The first circle is the set $\sigma_1$, the second is $\sigma_2$, and so on.)  We do this
according to the following provision: in order
to encircle $\delta_i$, if it is adjacent to some vertex $j$ which
has already been encircled by some $\sigma_{k}$, then $\sigma_i$ must
contain the $\sigma_{k}$ circle. %enclosing $\delta_i$ must also enclose $j$ and any circle previously made that contains it.  
Accordingly, the sets ``Left'' and ``Right'' keep track of how far to the left and right $\sigma_i$ must extend.
% (which will ensure that they form a valid tubing, which we discuss in Section \ref{Comb}).  
The result is  an unordered set $\sigma$ of $n{-}1$ encircled sets
$\sigma_1,\sigma_2,\ldots,\sigma_{n-1}$.  Figure 1 displays the beginning of an example of this encircling process for $n=11$.

%: Beginning- generating the signature as topographical map.
\begin{figure}[h]\label{tubingsignature}
\centerline{
\includegraphics[scale=1.8]{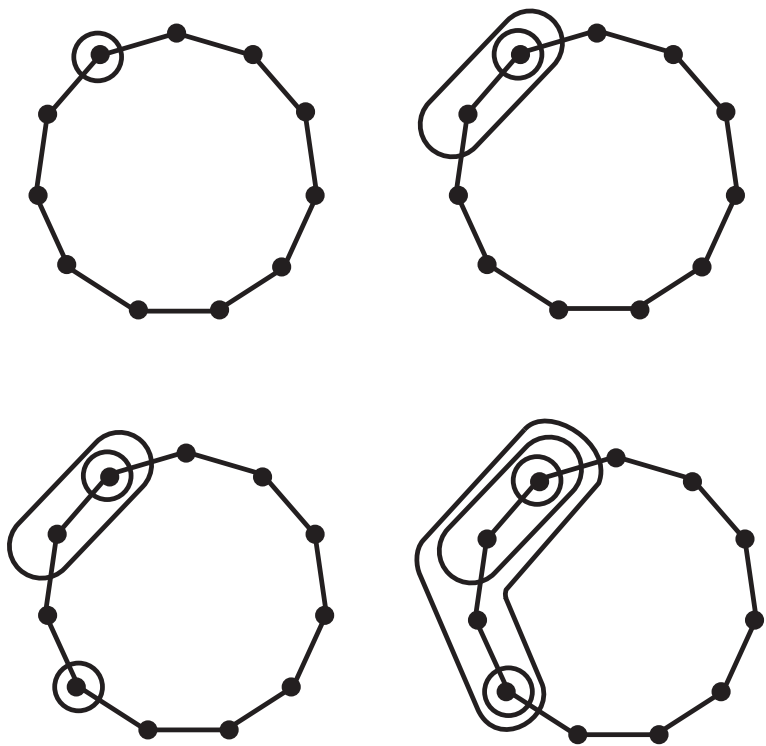}
}
\caption{Algorithm~\ref{cyctestalg} constructs a topographic map on the
  $n$-cycle by subsequently encircling vertices in order of decreasing size of the corresponding component of a data vector.  Displayed at the top are the formations of the first two components $\sigma _i$, and at the bottom are the third and fourth,
  of the signature for an example with $n=11$.}
\end{figure}

\begin{figure}[h]\label{tubingsmoothoncircle}
\centerline{
\includegraphics[scale=0.65]{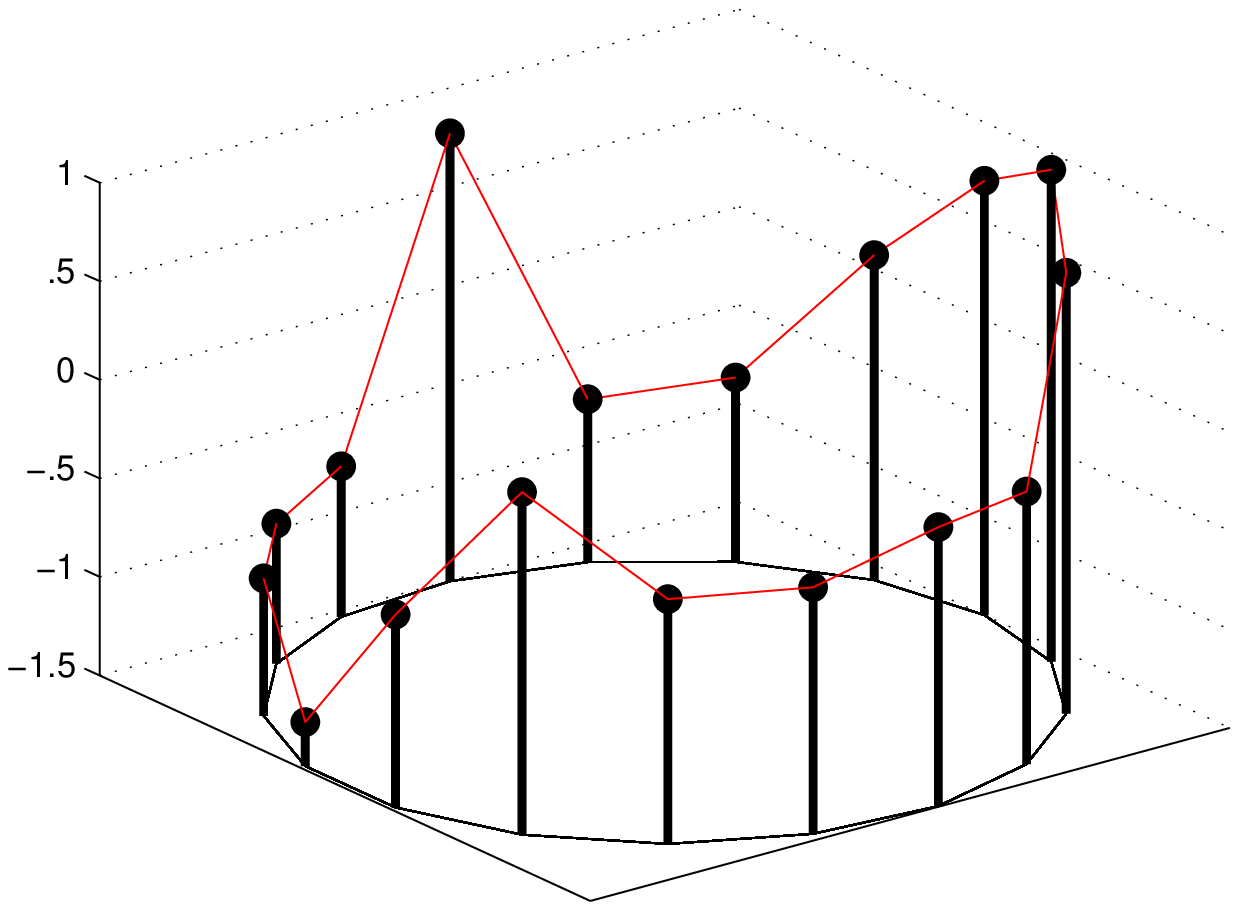}
\includegraphics[scale=0.65]{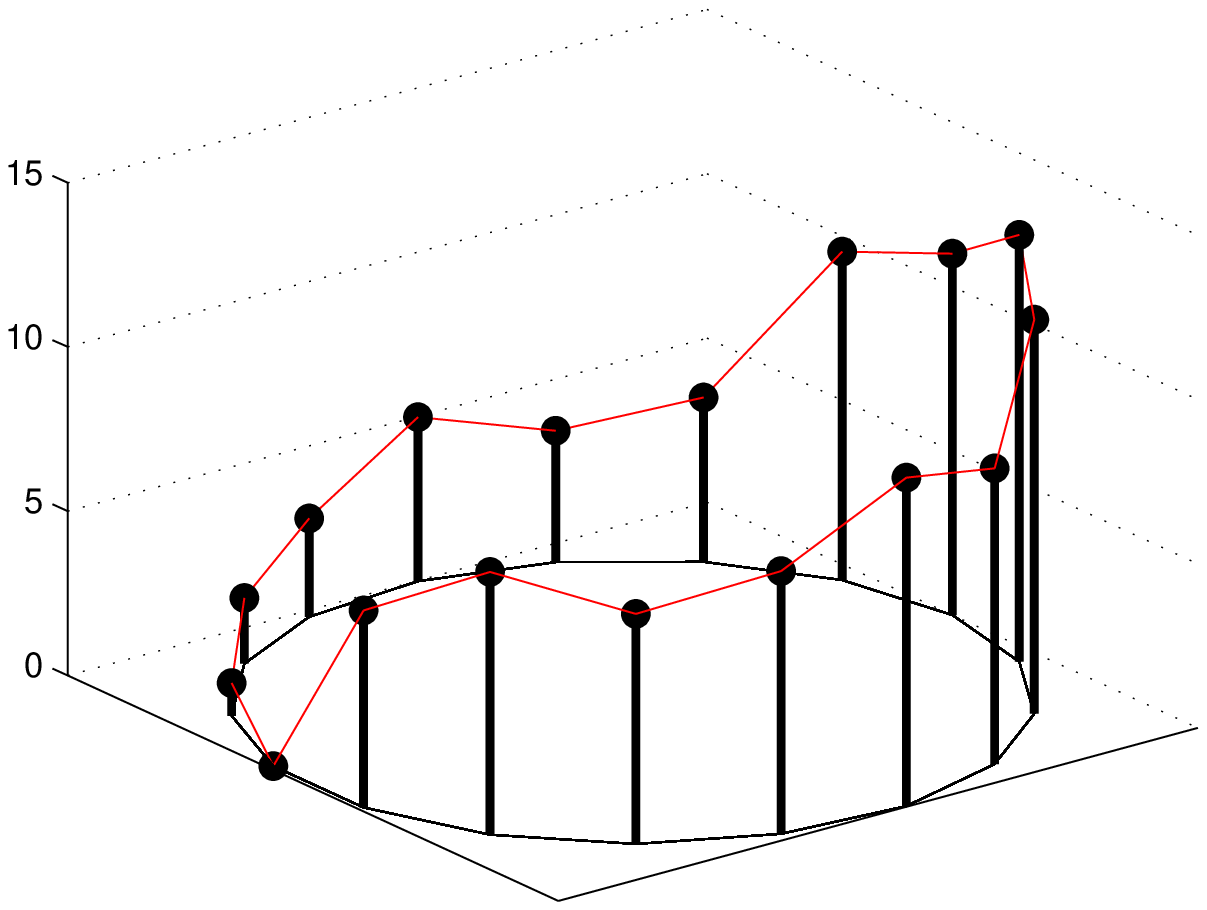}
}
\caption{The data $v$ (left) and the height vector $h(v)$ (right) for the gene {\tt Obox}.}
\end{figure}

We say that the height $h_i$ of the $i$-th vertex in the
topographic map for $v$ is the number of sets $\sigma_j$ which contain $i$.
We can identify the signature $\sigma$ with the
{\em height vector} $h = (h_1,h_2,\ldots,h_n)$, because
$\sigma$ can be recovered uniquely from the vector $h$.
The map $v \mapsto h(v)$ can be viewed 
as a {\em smoothing of the data}; see Figure \ref{tubingsmoothoncircle}.%This not a very precise statement.

\begin{remark}
The cyclohedron test applies when there
are no ties  $v_i = v_j$ in the data.
When ties occur, we examine all possible
permutations $\delta$ arising from small perturbations.
\end{remark}
%We explain the combinatorics and geometry of the cyclohedron test in Section \ref{Comb}.

%%%
%Applying the test to one
%data vector $v \in \R^n$ means computing its cyclic signature $\sigma(v)$
%and the probability $\,p(\sigma(v)) = \mathbf{c}(v)/n !$.
%Heuristically, if that probability is small then the data
%vector $v$ is likely to be cyclic.  That is, the significant data vectors are those with unusual (unlikely) topographical maps.

%Applying the cyclohedron test to  $N$
%data vectors $\,v^{(1)}, \ldots, v^{(N)}\,$
%in $\R^n$ means computing their permutation counts
%$\,\mathbf{c}(v^{(1)}),\ldots,\mathbf{c}(v^{(N)})$,
% and then ranking the $N$ data points according to these permutation counts.
% The highest ranked data are
%those  for which the permutation count $\,\mathbf{c}(v^{(i)})\,$ is smallest.  The assignment of p-values to such a ranking will be discussed in Section \ref{Sig}.
\begin{example}
\label{ex:obox}
In our analysis in Section 4, the
 number of microarray experiments is $n= 17$, and the number of
probesets (labels of the data vectors) is $N = 13,873$.
The probeset ranked first in Table~1 represents a gene named {\tt  Obox}.
Its data vector equals
\begin{eqnarray*} v \,\, = &
\bigl(0.738, 0.996,  0.705,  0.150, -0.566, -0.673,
0.774, -0.736, -0.788, \\ &  \, -0.802, -1.276, -0.521,
0.238, -0.258, -0.249, -0.084, -0.117 \bigr).
\end{eqnarray*}
The descent order permutation for this vector $v$ equals
$$\delta \,\, = \,\,
   \bigl(2, 7, 1, 3, 13, 4, 16, 17, 15, 14, 12, 5, 6, 8, 9, 10, 11 \bigr). $$
   The signature $\sigma$ is given by the unordered set
$\sigma_1 = \{2\}$, $\sigma_2 = \{7\}$,
$\sigma_3 = \{1, 2\}$,
$\sigma_4 = \{1, 2, 3\}$, etc.
The permutation count $\,\mathbf{c} \,=\, 480 \,$ is the product of
the three contributions made by $5$, $8$ and $12$, respectively,
when constructing
$\sigma_8 = \{1, 2, 3, 4, 16, 17\}$,
$\sigma_{10} = \{1, 2, 3, 4, 13, 14, 15, 16, 17\}$, and
$\sigma_{13} = \{1, 2, 3, 4, 5, 6, 7, 12, 13, 14, 15, 16, 17\}$.
When viewing $\sigma$ as a topographic map for the data $v$,
we obtain the height vector
$$ h(v) \,\, = \,\, \bigl(12, 13, 11, 10, 5, 4, 5, 3, 2, 1, 0, 6, 8, 7, 8, 10, 9 \bigr) .$$
Figure \ref{tubingsmoothoncircle} displays the data $v$ and the height vector $h(v)$
plotted around the circle. $\Box$
\end{example}

%:OBOX figure

\section{Significance testing}\label{Sig}

{\em Multiple hypothesis testing} is of concern in microarray experiments, because the number of hypotheses that are tested simultaneously is large. In our application, there are $N = 13,873$
null hypotheses. The hypotheses take the form ``the $r$ genes with the smallest counts {\bf c} arose by chance'',
for $r=1,2,\ldots,N$.   In this section, we explain how to assign p-values to these groups,
leading to a criterion for determining which hypotheses to reject.

\begin{figure}[ht] \label{c17pdf}
\centerline{
\includegraphics{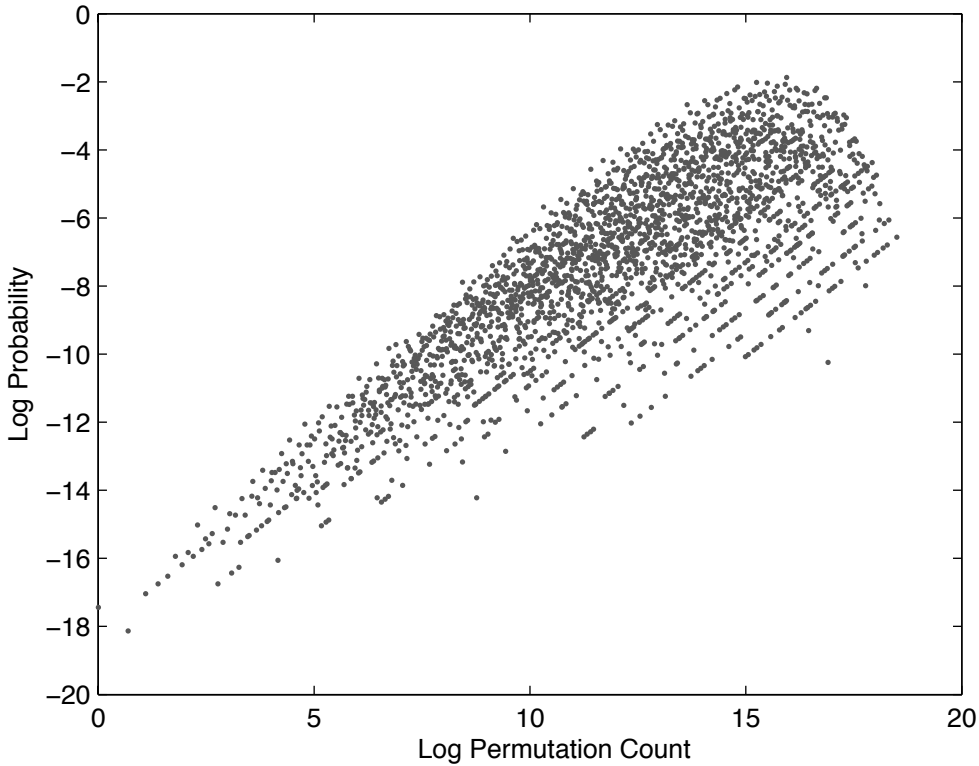} 
}
\caption{The probability distribution function of $\mathbf{c}$ for $n=17$.}
\end{figure}

Applying the cyclohedron test to $N$
data vectors $v^{(1)}, \ldots, v^{(N)}$
in $\R^n$ means computing their permutation counts
$\,\mathbf{c}(v^{(1)}),\ldots,\mathbf{c}(v^{(N)})\,$. The highest ranked data are
those  for which $\,\mathbf{c}(v^{(i)})\,$ is smallest.  Under the null hypothesis, the probability distribution on $\R^n$ of each data vector $v^{(i)}$
 induces the uniform distribution $U$ on the $n! $ permutations $\delta$. 
Viewed as a random variable, the permutation count $\mathbf{c}$ has probability distribution function
\begin{equation}
\label{pdf}
 P_{\mathbf{c}}  \,: \, \im(\mathbf{c}) \rightarrow [0,1]\, , \,\,\,\,
 \gamma \,\mapsto \, \Pr _{ U} (\mathbf{c}(\delta)= \gamma ).
 \end{equation}
 Here $\, \im(\mathbf{c}) \, = \,
 \{ \gamma_1 < \gamma_2 < \cdots < \gamma_{s_n} \}\,$ is the
 set of all positive integers that arise as permutation counts
 $\mathbf{c}(\sigma) $ for some $\sigma \in \mathcal{C}_n$.  
The probability distribution function $ P_{\mathbf{c}}$ is displayed in Figure \ref{c17pdf} for $n=17$,
which will be the number of time points in Section 4.
 \vskip 0.2in %for figure

We now fix two integers $\,1 \leq r \leq N$. The
 {\em order statistic} $\,{\bf C}_{(r)}\,$
 %corresponding to the probability distribution $\,P_{\mathbf{c}}\,$ 
 is the function
$\, \R^{N \times n} \rightarrow \im(\mathbf{c})\,$
 which  takes any list of $N$ data vectors $\, V = (v^{(1)}, \ldots,v^{(N)})\,$
 and returns the $r^{th}$ smallest value among the permutation counts
 $\,\mathbf{c}(v^{(1)}), \dots, \mathbf{c}(v^{(N)})$.
 Recall that under the null hypothesis, each $v^{(i)}$ has a distribution on $\R^n$ which
 induces the uniform distribution on permutations $\delta$.  Further let us assume that 
 %under the null data generating distribution, 
 the data vectors $v^{(i)}$ are independent.  This induces a joint distribution $Q_0$ on the vector $\left( \,\mathbf{c}(v^{(1)}), \dots, \mathbf{c}(v^{(N)})~\right)\in \R^{N \times n}  $ of counts.  In this framework, we  view the order statistic $\,{\bf C}_{(r)}\,$ as a random variable with distribution
 $$ F_{(r)}:\im(\mathbf{c}) \rightarrow [0,1],\,
 \gamma \,\mapsto \, \Pr_{Q_0}( {\bf C}_{(r)} = \gamma). $$
In other words, $F_{(r)}(\gamma)$ is the probability that
 the $r$-th smallest value among
 the permutation counts $\,\mathbf{c}(v^{(1)}), \dots, \mathbf{c}(v^{(N)})\,$
 of $N$ random data vectors equals $\gamma$.
The function $F_{(r)}(\gamma)$ depends only on $n, N$ and $r$.
Its efficient computation is explained In Section \ref{Comput}.

\begin{defn} {\bf (p-value) } Suppose we apply the cyclohedron test
to $N$ data vectors in $\R^n$, and the data vector whose
permutation count is the $r$-th smallest has permutation count
 $\gamma_k$. Then the collective {\em p-value} of the group of $r$ highest ranked data vectors is
\begin{equation}
\label{pvalue}
\Pr_{Q_0} ({\bf C}_{(r)} \leq \gamma_k) \quad = \quad
F_{(r)}(\gamma_1) +
F_{(r)}(\gamma_2) + \cdots +
F_{(r)}(\gamma_k) .
\end{equation}
The p-value (\ref{pvalue}) is the probability that
the $r$-th order statistic for random data under the null would be 
less or equal to the value of the $r$-th order statistic for the observed data.
\end{defn}

We now offer some remarks regarding how our multiple testing procedure differs from those typically used when the number of hypotheses is large (say, in the thousands).  To analyze gene
expression data, it is often appropriate to employ a joint null distribution $Q$ that allows for dependencies among the genes.  These dependencies are unknown, so an estimate of the null distribution of the test statistics (for the cyclohedron test, the vector of counts) is made from bootstrap samples of the data; two such estimates are the null shift and scale \cite{MTP} and null quantile \cite{nullQ} distributions.  Further, there is typically one null hypothesis per gene, and p-values are assigned to control some general Type-I error rate.  However, our chosen joint null distribution $Q_0$ is simple and can be computed exactly.  In addition, this test was motivated by the exploratory data analysis described in the next section.  Having a powerful procedure was not critical; rather, the aim was to identify groups of top genes for further biological testing.  In other settings, however, different choices of joint null distribution or of multiple testing procedure (such as those in \cite{MTP}) can improve power.  

\section{Application to mouse microarray data} \label{Mouse}

We applied the cyclohedron test to microarray data from recent work that investigated the mouse segmentation clock \cite{Science}.   Dequ\'{e}ant {\em et al.} took 17 expression measurements from mouse presomitic mesoderm on Affymetrix MOE430A arrays.
By independently measuring the expression of the gene Lunatic Fringe
({\tt Lfng}) which is known to be periodic within the somitogenesis
cycle of embryonic development,
Dequ\'{e}ant {\em et al.} ordered the 17
experiments within the cycle.
Each array consisted of over $22,000$ probesets, however
we restrict the analysis to a subset of $13,873$ probesets by removing
genes whose expressions are deemed ``absent'' across the experiments by
Affymetrix standards.  In other words, the data consisted of 13,873 data
vectors $v$, each of which was the expression level of one gene (divided
by the mean across experiments and transformed to $\log_2$).  We then
applied the cyclohedron test to these data.  
We were interested in those genes whose counts $\mathbf{c}(v)$ were
small.  Accordingly, we ranked
 the genes by their counts; Table 1 presents the first 32 genes.  Table
 2 lists the significance of top groups of genes.  For example, the
 first 32 genes collectively have a p-value of 0.081, which suggests
 that these 32 genes are of interest.
\begin{table}
\begin{center}
 \begin{tabular}{ | c || c | c | c | c | c |}
   \hline
    Rank  & ProbeSet & Gene Name & Gene Description & Count  \\ \hline
1 & 1456017\_x & Obox1 &  similar to oocyte specific gene & 480 \\
2 & 1452041 & Klhl26 & kelch-like 26 (Drosophila) & 1440 \\
3 & 1418593 & Taf6 & TAF6 RNA polymerase II & 1560 \\
4       &       1417985     &       Nrarp   &       Notch-regulated ankyrin repeat protein &  1950 \\
5       &       1436845     &       Axin2   &       axin2  &   2240 \\
5       &       1436343     &       Chd4    &       chromodomain helicase DNA binding protein    & 2240 \\
7       &       1426267     &       Zbtb8os &       zinc finger and BTB domain      & 2310 \\
8       &       1420360     &       Dkk1    &       dickkopf homolog 1 (Xenopus laevis)     & 2520 \\
9       &       1449643\_s  &       Btf3    &       basic transcription factor 3    & 2772 \\
10      &       1417399     &       Gas6    &       growth arrest specific 6        & 2800 \\
11      &       1418102     &       Hes1    &       hairy and enhancer of split 1 (Drosophila)      & 3120 \\
12      &       1448799\_s  &       Mrps12  &       mitochondrial ribosomal protein S12     & 3150 \\
13      &       1418729     &       Star    &       steroidogenic acute regulatory protein  & 3600 \\
14      &       1425424     &       MGC7817     &       hypothetical protein LOC620031  & 3850 \\
15      &       1455740     &       Hnrpa1  &       heterogeneous nuclear ribonucleoprotein      & 4004 \\
16      &       1450204\_a  &       Mynn    &       myoneurin       & 4928 \\
17      &       1449120\_a  &       Pcm1    &       pericentriolar material 1       & 6006 \\
18      &       1423106     &       Ube2b   &       ubiquitin-conjugating enzyme E2B        & 6720 \\
19      &       1420386     &       Seh1l   &       SEH1-like (S. cerevisiae)        & 6825 \\
20      &       1456380\_x  &       Cnn3    &       calponin 3, acidic      & 8008 \\
21      &       1419438     &       Sim2    &       single-minded homolog 2 (Drosophila)    & 8640 \\
22      &       1426524     &       Gnpda2  &       glucosamine-6-phosphate deaminase 2     & 9009 \\
23      &       1438557\_x  &       Dnpep   &       aspartyl aminopeptidase & 9450 \\
24      &       1454904     &       Mtm1    &       X-linked myotubular myopathy gene 1     & 10500 \\
25      &       1448951     &       Tnfrsf1b        &       tumor necrosis factor receptor
superfamily     & 10530 \\
25      &       1433952     &       Tufm     &       Tu translation elongation factor  & 10530 \\
27      &       1422327\_s  &       $^{G6pd2/}_{G6pdx}$ &       glucose-6-phosphate
dehydrogenase 2 & 10725 \\
28      &       1416295\_a  &       Il2rg   &       interleukin 2 receptor, gamma chain     & 10920 \\
29      &       1417316     &       Them2   &       thioesterase superfamily member 2       & 11025 \\
30      &       1450242     &       Tlr5    &       toll-like receptor 5    & 11232 \\
31      &       1449164     &       Cd68    &       CD68 antigen    & 11340 \\
32      &       1418337     &       Rpia    &       ribose 5-phosphate isomerase A  &11760 \\

\vdots & \vdots & \vdots & \vdots & \vdots \\
\hline
   \hline
 \end{tabular}
\caption{The 32 genes ranked highest by the cyclohedron test. Gene descriptions are
derived from those provided by Affymetrix. The suffix ``\_at'' was removed from each ProbeSet ID.}
\end{center}
\end{table}
At this point we  recall the definition of a p-value which
 was given in equation (\ref{pvalue}).
The p-value of the rank-$1$ gene
{\tt Obox1} is the probability under the null hypothesis that the top-ranked permutation count
is less than or equal to $480$, while the p-value of the first $16$ genes (the number
$0.008$ in Table 2) is the probability that the gene ranked $16$ has permutation
count less than or equal to $4928$.
It is important to emphasize that the p-values do not reveal  the
significance of any individual gene, but rather of a collection  of
genes. For example, the top $19 $ genes having a collective p-value of $0.046$
means this: the probability that the first 19 genes would collectively
all have permutation count at most $6825$ under the null distribution is $0.046$. In other words, the
group as a whole is significant. However, we determine whether any
individual gene in that group is significant.  
For example, there is no significance to the fact that
{\tt Obox1} is ranked first. While it  appears to be the most periodic pattern in
the data by our analysis,
that could have happened by chance
(p-value 0.279).
A natural cutoff value is
to look at the first
32 genes because collectively they have a p-value of
$0.081 $ (the next ten p-values are between 0.13 and 0.30). Note that analyses of microarray data have this property, that Type-1 errors are all but guaranteed due to the large number of genes (and thus the large number of hypotheses) that are tested.  Our computations were performed with the statistical software {\tt R} \cite{R}, using the implementation described in the Appendix.

Dequ\'{e}ant {\em et al.} performed significance testing according to a
Lomb-Scargle analysis, and then based on gene expression profile
clustering, they identified genes belonging to three
pathways Notch/FGF and Wnt that are involved with somitogenesis.  There
are genes that are deemed interesting by both the analysis of
Dequ\'{e}ant {\em et al.} and the cyclohedron test.  For example, {\tt Axin2} is
ranked highly by the Lomb-Scargle (rank 6) and the cyclohedron test
(rank 5).  In addition, {\tt nrarp} (rank 4 according to the cyclohedron test)
is ranked poorly by Lomb-Scargle (rank 482), although it belongs to the
Notch pathway and its gene expression clusters accordingly.  Finally,
there are novel genes such as {\tt Obox} (rank 1 by the cyclohedron test, but
not known to be related to somitogenesis) that require further
investigation.   This suggests that to find periodic gene expression, it
is beneficial to apply many methods, including Lomb-Scargle, clustering,
and the cyclohedron test.  Doing so enables us to find genes overlooked
by each method, as well as to confirm findings of other tests.  In other
words, the findings of each method complement those of others by identifying candidate genes for knockout experiments.  The forthcoming paper  \cite{compare} 
will compare various methods, including Lomb-Scargle, up-down analysis,
 and the cyclohedron test, for identifying cyclic genes from this data set.

\begin{table}
\begin{center}
 \begin{tabular}{ | l || c | c | c | c | c | c | c | c |}
   \hline
    Group & 1..1 & 1..2 & 1..3 & 1..4 & 1..5 & 1..6 & 1..7& 1..8  \\
 \hline
p-value & 0.279 & 0.458 & 0.244 & 0.204 & 0.064 & 0.064 & 0.031 &
0.020  \\  \hline
 \end{tabular}

   \begin{tabular}{ | l || c | c | c | c | c | c | c | c |}
   \hline
    Group & 1..9 & 1..10 & 1..11 & 1..12 & 1..13 & 1..14 & 1..15&
1..16  \\ \hline
p-value & 0.014 & 0.005 & 0.005 & 0.002 & 0.003 & 0.003 & 0.002 &
0.008  \\
 \hline
 \end{tabular}

     \begin{tabular}{ | l || c | c | c | c | c | c | c | c |}
   \hline
    Group & 1..17 & 1..18 & 1..19 & 1..20 & 1..21 & 1..22 & 1..23&
1..24  \\ \hline
p-value & 0.047 & 0.069 & 0.046 & 0.139 & 0.165 & 0.173 & 0.195 &
0.312  \\
 \hline
 \end{tabular}

       \begin{tabular}{ | l || c | c | c | c | c | c | c | c |}
   \hline
    Group & 1..25 & 1..26 & 1..27 & 1..28 & 1..29 & 1..30 & 1..31&
1..32  \\ \hline
p-value & 0.192 & 0.192 & 0.168 & 0.159 & 0.118 & 0.096 & 0.075 &
0.081  \\
 \hline
 \end{tabular}

\end{center}
\caption{Significance of top-ranked groups of genes.  For example, the first 32 genes have a collective p-value of 0.081.}
\end{table}

In conclusion, we remark that, although microarray expression analyses
are frequently criticized due to the noise in individual measurements,
the massively parallel nature of the experiments provide the
possibility for finding groups of significant genes. Indeed, we confirm
this in our 
analysis of the Dequ\'{e}ant {\em et al.} experiments \cite{Science},
in which we are unable to confirm whether any individual gene is statistically significant, yet we can identify a group of genes that collectively
are significant. The biological significance 
of individual genes can be determined by further
targeted experimental validation. 

% Issue: ties in data 

\section{Combinatorics of the cyclohedron test}\label{Comb}

We now describe the combinatorics and geometry
behind our test.  First, the set $\mathcal{C}_n$ of cyclic signatures is in
natural bijection with the vertices of a certain convex polytope.
The $n$-cycle has $n(n{-}1)$ connected induced proper subgraphs, namely,
the {\em cyclic segments} of the form $\, S = \{i,i+1,\ldots,i+k\} $. Here $k <n{-}1$, and the indices
are understood modulo $n$. The {\em cyclohedron vertex} of
a data vector $v \in \R^n$ is the vector $\tau(v) \in \N^n $ whose
$i$-th coordinate $\tau(v)_i$ is the number of cyclic segments
$S$ containing $i$ such that $v_i = {\rm min} \{v_s : s \in S\}$.
The {\em cyclohedron} $C_n$ is the convex hull in $\R^n$ of
all the cyclohedron vertices $\,\tau(v)\,$ where $v$ ranges over $\R^n$.
For $n=4$ and the data vector $v = (0.49, 5.73, 4.01, 2.67)$, we  have
$\tau(v) = (6, 1, 2, 3)$,  while for 
$v' = (0.49, 5.73, 2.67,4.01)$ we have
$\tau(v') = (6, 1, 4, 1)$.  For example, $\tau({v})_3=2$ because $v_3=4.01$ is minimal in $S^1=\{3\}$ and $S^2= \{2,3\}$. 

Two vectors in $\R^4$ share
the same signature $\sigma = \{\sigma_1,\sigma_2,\sigma_3\}$
if and only if they are mapped to the same cyclohedron vertex $\tau$.
The convex hull of all cyclohedron vertices $\tau (v)$ is the $3$-dimensional cyclohedron $C_4$.
This is  a simple polytope with $20$ vertices, $30$ edges and $12$ facets
(for the $12$ cyclic segments). It is depicted in Figure 3.  Vertices in the figure, incident to a `double' edge indicate
signatures $\sigma$ with $\mathbf{c}(\sigma)=2$.
 Thus the set $\mathcal{C}_4$ of all
signatures has $20$ elements, one for each vertex of $C_4$.

\begin{figure} \label{cyclohedron}
\[
  \begin{xy}<2.5cm,0cm>:
%Top pentagon- its five vertices
(2.3,6.4)="2143" *+!DR{2143}*{\bullet};
(2.75,6.7)="1243" *+!DR{1243}*{\bullet};
(3.3,6.7)="1423" *+!DL{1423}*{\bullet};
(3.8,6.4)="4123" *+!DL{4123}*{\bullet};
(3.05,5.9)="^2_413" *+!UR{^2_413}*{\bullet};
%Top- its five edges
"2143";"1243"**@{-}; %1 indep 2
"1243";"1423"**@{-}; 
"1423";"4123"**@{-}; 
"4123";"^2_413"**@{.}; 
"^2_413";"2143"**@{.}; 
%Left pentagon- its five vertices
(1.9,4.7)="3214" *+!L{3214}*{\bullet};
(1.4,5)="2314" *+!DR{2314}*{\bullet};
(1.4,5.4)="2134" *+!DR{2134}*{\bullet};
(1.9,5.6)="1234" *+!L{1234}*{\bullet};
(2.3,5.2)="^1_324" *+!UL{^1_324}*{\bullet};
%Left- its five edges
"3214";"2314"**@{-}; 
"2314";"2134"**@{-}; 
"2134";"1234"**@{-}; 
"1234";"^1_324"**@{-}; 
"^1_324";"3214"**@{-}; 
%Right pentagon- its five vertices
(4.2,5.6)="1432" *+!R{1432}*{\bullet};
(4.6,5.4)="4132" *+!DL{4132}*{\bullet};
(4.6,4.9)="4312" *+!UL{4312}*{\bullet};
(4.2,4.7)="3412" *+!R{3412}*{\bullet};
(3.75,5.2)="^1_342" *+!UR{^1_342}*{\bullet};
%Right- its five edges
"1432";"4132"**@{-}; 
"4132";"4312"**@{-}; 
"4312";"3412"**@{-}; 
"3412";"^1_342"**@{-}; 
"^1_342";"1432"**@{-}; 
%Bottom pentagon- its five vertices  
(3.8,3.9)="4321" *+!UL{4321}*{\bullet};
(3.3,3.7)="3421" *+!UL{3421}*{\bullet};
(2.75,3.7)="3241" *+!UR{3241}*{\bullet};
(2.3,3.9)="2341" *+!UR{2341}*{\bullet};
(3.05,4.45)="^2_431" *+!DR{^2_431}*{\bullet};
%Bottom- its five edges
"4321";"3421"**@{-}; 
"3421";"3241"**@{-}; 
"3241";"2341"**@{-}; 
"2341";"^2_431"**@{.}; 
"^2_431";"4321"**@{.}; 
%Upper Left Square - two edges
"2143";"2134"**@{-}; 
"1243";"1234"**@{-}; 
%Upper Right Square - two edges
"1423";"1432"**@{-}; 
"4123";"4132"**@{-}; 
%Lower Left Square - two edges
"2314";"2341"**@{-}; 
"3214";"3241"**@{-}; 
%Lower Right Square - two edges
"3412";"3421"**@{-}; 
"4312";"4321"**@{-}; 
%Doubled Edge in the back
"^2_413";"^2_431" ** @{:};
%Doubled Edge in the front
"^1_324";"^1_342" ** @{=};
   \end{xy}
\]
\caption{The cyclohedron $C_4$; its vertices correspond to the distinct signatures for $n=4$.  Following the notation of \cite{semigraphoid}, 
the string $^1_324$ labels the cyclohedron vertex of data vectors whose descent permutation is $1324$ or $3124$.}
\end{figure}
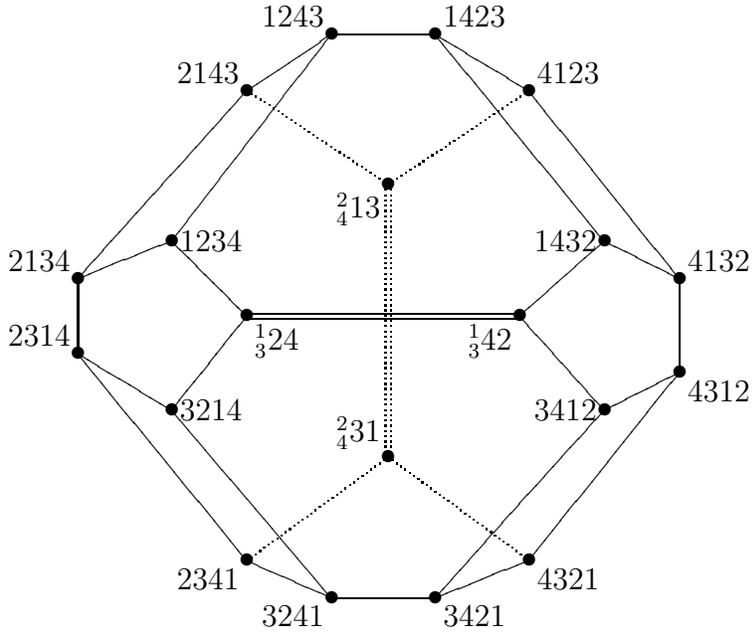

The following theorem summarizes what is known about the cyclohedron.
It is extracted from 
\cite{Fomin, Hohlweg, Markl}.

\begin{thm} \label{propcyclo}
The cyclohedron $C_n$ is an $(n{-}1)$-dimensional polytope.
It is the solution set in $\R^n$ of the following system of one
linear equation and $n(n-1)$ linear inequalities:
\begin{equation}
\label{sys1}
 x_1 + x_2 + \cdots + x_n  \,\, = \,\, n(n-1), \qquad   \qquad
 \end{equation}
 \begin{equation}
 \label{sys2}
 \qquad \qquad
 \sum_{s \in S} x_s \, \geq\,  \binom{|S|+1}{2}
\quad \hbox{for each cyclic segment $S$}.
\end{equation}
The cyclohedron $C_n$ is simple, i.e.~each vertex lies on precisely $n{-}1$ facets.
Each inequality (\ref{sys2}) defines a facet.
The total number of vertices equals $\binom{2n-2}{n-1}$.
More generally,
the number $f_i $ of $i$-dimensional faces of the
cyclohedron is given by the generating function
\begin{equation}
\label{nirit}
\sum_{i=0}^{n-1} f_i \cdot z^i \quad = \quad
\sum_{k=0}^{n-1} \binom{n-1}{k}^2 \cdot (z+1)^k.
\end{equation}
\end{thm}

Algorithm \ref{cyctestalg} is a greedy method for
linear programming on the cyclohedron $C_n$.
Indeed, computing the cyclohedron vertex $\tau(v)$ of a data
vector $v = (v_1 ,\ldots,v_n)$ is equivalent to the
linear program of minimizing $\,\sum_{i=1}^n v_i x_i\,$
subject to the constraints (\ref{sys1}) and (\ref{sys2}).
The optimal vertex of that linear program on $C_n$
is precisely the vector $\,x = \tau(v)$.

Given the
linear functional $\, \sum_{i=1}^n v_i x_i \,$ to be minimized,
Algorithm \ref{cyctestalg} generates a collection
$\,\sigma = \{\sigma_1, \sigma_2, \ldots, \sigma_{n-1}\}\,$ of
subsets of $\{1,2,\ldots,n\}$. These sets
$\,S = \sigma_i\,$ are cyclic segments, and they indicate
which $n{-}1$ inequalities (\ref{sys2}) are tight at the
optimal vertex $x = \tau(v)$ of $C_n$.
This implies that $\tau(v)$ can be recovered
from $\sigma(v)$ and vice versa:

\begin{cor} \label{sigmatau}
The cyclohedron vertex $\tau(v)$ of any data vector $v \in \R^n$ can be obtained from the
signature $\,\sigma(v) = \{\sigma_1,\sigma_2,\ldots,\sigma_{n-1}\}\,$
by solving the
linear system  of equations
{\em
\begin{equation}
\label{oneless}
 \hbox{(\ref{sys1})} \quad \hbox{and} \quad
 \sum_{s \in \sigma_i} x_s \,  = \,  \binom{|\sigma_i|+1}{2}
\quad \hbox{for}\,\,  i = 1,2,\ldots,n{-} 1.
\end{equation}}
Conversely, the signature $\sigma(v)$ is recovered
from the  vertex $\tau(v)$ by substituting
$ x = \tau(v)$ into the inequalities (\ref{sys2}) and
collecting all index sets $S$ for which equality holds.
\end{cor}

 In light of Corollary \ref{sigmatau},
we henceforth shall identify signatures $\sigma \in \mathcal{C}_n$
with their corresponding vertices $\tau$ of the cyclohedron $C_n$.
We note that the solution $\tau$ to (\ref{oneless}) can be read off
easily within Algorithm \ref{cyctestalg}.
It always holds that $\, \tau_{\delta_n} \,\, := \,\,\binom{n}{2}$,
and the other $n{-}1$ coordinates are obtained by adding one line at the end of the main $i$ loop:

\medskip

\qquad {\bf Output} $\,\, \tau_{\delta_i} \,\, = \,\,(|{\rm Left}|+1) \cdot (|{\rm Right}| + 1)$.

\bigskip

Two data vectors $v$ and $v'$ are {\em cyclically equivalent}
if and only if $\,\sigma(v) = \sigma(v')\,$, i.e., if and only if
the linear functionals corresponding to $v$ and $v'$ are minimized
at the same vertex $\tau(v) = \tau(v')$ of the cyclohedron $C_n$.
The cyclic equivalence classes are the normal cones at the
 vertices of $C_n$.   They are specified by the inequalities
 $\,v_{_{\delta_i}} < v_{_{\delta_k}}\,$
 for all inclusions $\sigma_k \subset \sigma_j$ in $\sigma(v)$.
 Since $C_n$ is simple, $n{-}1$ inequalities suffice, and these
 can be generated by augmenting Algorithm \ref{cyctestalg}, again at the end of the main $i$ loop, as follows:

\smallskip

\qquad if $\,{\rm Right} \neq \emptyset \,$ or
 $\,{\rm Left} \neq \emptyset \,$ then {\bf output} $\,v_{\delta_i} < v_{\delta_k}$.

\smallskip

\noindent The generated inequalities permit the study of confidence regions for
 the cyclohedron test.

 \begin{example} \label{ex:obox2}
Fix $n=17$ and let $v$ be the data vector
for the {\tt Obox} gene in Example \ref{ex:obox}.
The augmented Algorithm \ref{cyctestalg} reveals that the
cyclic equivalence class of $v$ is given by
\begin{eqnarray*}
& v_{11} < v_{10} < v_9 < v_8 < v_6 < v_5 < v_{12}  < v_{14} <
    v_{15} < v_{17} < v_4 < v_3 < v_1 < v_2 \\
& \,\,\, \hbox{and} \quad  v_6 < v_7
 \quad \hbox{and} \quad  v_{17} < v_{16}
 \quad \hbox{and} \quad  v_{14} < v_{13}.
 \end{eqnarray*}
These inequalities specify the normal cone at the vertex
$$ \tau(v) \,=\, \bigl(2, 1, 3, 4, 11, 24, 1, 14, 15, 16, 136, 10, 1, 16, 7, 1, 10  \bigr) $$
of the $16$-dimensional  cyclohedron $C_{17}$.
Recall that the possible signatures for data with $n=17$
are (in bijection with) the vertices of $C_{17}$, and their total number equals
$$ |\mathcal{C}_{17}| \,\, = \,\,
\binom{2\cdot 17 -2}{17-1} \,\, = \,\, \binom{32}{16}\,\, = \,\, 601,080,390 . $$
Among all these signatures,
the vertex $\tau(v)$ is of interest because
the probability that a random linear functional attains its minimum over
$C_{17}$ at that vertex is rather small:
$$ p(v) \,\, = \,\, \mathbf{c}(v)/n ! \,\,  =\,\, 480/17 ! \,\, = \,\,
1.35 \cdot 10^{-12}. $$
The results of our analysis for the full data set were presented in Section 4. \qed
 \end{example}

 The theory of graph associahedra
also offers the following combinatorial characterization of the
possible outputs of Algorithm \ref{cyctestalg}.
A collection $\,\{\sigma_1,\sigma_2, \ldots \}\,$
of cyclic segments is called a {\em tubing}
of the $n$-cycle if any two elements satisfy the following property: either
$\sigma_i \subset \sigma_j$, or $\sigma_j \subset \sigma_i$, or
$\sigma_i$ and $\sigma_j$ are disjoint and no
node in $\sigma_i$ is adjacent to a node in $\sigma_j$
Each maximal tubing has the same number of elements,
namely $n {-} 1$, and the maximal tubings are precisely
the signatures generated by Algorithm \ref{cyctestalg}.
The simplicial complex of all tubings is dual to the
face poset of the simple polytope $C_n$.
Analogous statements hold for the face poset of the
graph associahedron of any graph $G$ with vertex set
$\{1,2,\ldots,n\}$. The cyclohedron $C_n$
is the special case when $G$ is the $n$-cycle.

We propose that
rank tests  which are associated with graphs $G$ in this manner
be called {\em topographical models}. This is motivated by their
relationship with graphical models (Markov random fields) which was developed
in \cite{GRT}. Our cyclohedron vertex map $\tau$, for
$G$ the $n$-cycle, was denoted $\,\tau^*_{{\mathcal K}(G)}$
in \cite[\S 5]{GRT}. We believe that topographical models
for graphs $G$ other than the $n$-cycle will be useful
for wide range of statistical problems concerning data with
an underlying graphical structure.

\section{Null distribution of the counts and order statistics}\label{Comput}

We next compute two probability distribution functions, that of the random variable ${\mathbf{c}}$ and of its order statistics.
In the first part of this section we introduce a generating function
 that represents the distribution $P_{\mathbf{c}}$ of $\mathbf{c}$ under the null distribution.
 This is applied in the second part to derive the order statistics
of  $\,P_{\mathbf{c}}\,$ and a formula for
 computing the collective p-values (\ref{pvalue}) {\em exactly}.
\
Recall that the set $\mathcal{C}_n$ of signatures equals the set of
maximal tubings or vertices of the
cyclohedron $C_n$.  For each $\sigma \in \mathcal{C}_n$,
the quantity $ \,\mathbf{c}(\sigma) = p(\sigma) \cdot n !\,$ is the
number of permutations $\delta$ which map to $\tau$.
See Algorithm \ref{cyctestalg} and Proposition \ref{propscore}.

We define the {\em count generating function} for the cyclohedron test to be the polynomial
$$ \Gamma_{n} (t )\,\,\,\,\,:=  \,\,\, \sum_{\sigma \in \mathcal{C}_n}   t^{\mathbf{c}(\sigma)}.$$
By Theorem \ref{propcyclo},
this polynomial gives a refinement of the central binomial coefficient:
$$ \Gamma_n(1) \,\,=\,\, | {\rm Vert}(C_n)| \,\, = \,\,  \binom{2n-2}{n-1} .$$
Similarly, the first derivative $\,\Gamma'_n(t) = \frac{d}{dt} \Gamma_n(t)\,$
gives a refined count of the permutations:
$$ \Gamma'_n(1) \,\, = \,\, | \Sigma_n | \,\, = \,\, n ! . $$
We list the first few non-trivial instances of the count generating function:
\begin{eqnarray*}
\Gamma_4(t) & = & 4 t^2+16 t, \\
\Gamma_5(t) & = & 20 t^3+10 t^2+40 t, \\
\Gamma_6(t) & = & 12 t^8+24 t^6+48 t^4+48 t^3+24 t^2+96 t, \\
\Gamma_7(t) & =  & 28 t^{20}+56 t^{15}+140 t^{10}+28 t^8+56 t^6+112 t^5 \! + \! 112 t^4 \! + \!112 t^3
\! + \! 56 t^2 \!+ \! 224 t, \\
\Gamma_8(t) & = & 8 t^{80}+32 t^{48}+128 t^{45}+64 t^{40}+64 t^{36}+64 t^{30}+
\cdots + 256 t^3 \! + \! 128 t^2 \! + \! 512 t, \\
\Gamma_9(t) & =  & 72 t^{210}+72 t^{168}+108 t^{140}+144 t^{126}+432 t^{105}+\cdots
+ \! 576 t^3 \! + \! 288 t^2 \! + \!1152 t.
\end{eqnarray*}
The count generation function encodes the probability distribution function of $\mathbf{c}$:
\begin{remark} \label{rem:genf}
The probability $\,P_{\mathbf{c}}(\gamma) \,$ is the
coefficient of $\,t^\gamma \,$ in the polynomial $\, (t/n !) \cdot \Gamma'_n(t)$.
\end{remark}

\begin{example} \label{ex:seven}
Consider the case $n=7$. The $s_7  = 10$ possible permutation counts are
$$ \, \im(\mathbf{c}) \, = \, \bigl\{1,2,3,4,5,6,8,10,15,20 \bigr\}. $$
The probability for each of these counts to be observed
is the corresponding coefficient in
$$ \sum_{\gamma \in \im(\mathbf{c})} \! P_{\mathbf{c}}(\gamma) \cdot t^\gamma \quad  = \quad
\frac{t}{5040} \cdot \Gamma'_7(t) \quad = \quad
 \frac{1}{9} t^{20}  + \frac{1}{6} t^{15} + \frac{5}{18} t^{10} + \cdots +
\frac{1}{45} t^2 + \frac{2}{45} t. $$
For instance, the cyclohedron $C_6$ has $56$ vertices $\sigma$ with $\mathbf{c}(\sigma) = 15$,
and this accounts for $\,56 \cdot 15 = 840\,$ of the $5040$ permutations $\delta$ in $\Sigma_7$.
Thus the probability that a random data vector $v \in \R^7$ has permutation count
$\,\mathbf{c}(v) = 15\,$ is equal to $\,P_{\mathbf{c}}(15) =  840/5040 = 1/6$. \qed
\end{example}

We now describe a formula for computing the count generating function.
Let $\mathcal{T}_{m}$ denote the set of unlabeled rooted trees with
$m$ nodes, where each node has at most two children.
The number of these trees is the
{\em Wedderburn-Etherington number}, denoted by $\, t_m := |\mathcal{T}_m|$.
Starting with $t_0 = 1$, the Wedderburn-Etherington numbers are
$$ 1, 1, 2, 3, 6, 11, 23, 46, 98, 207, 451, 983, 2179, 4850, 10905, 24631,
56011, 127912 ,\ldots $$
and they can be computed by the following recursion:
\begin{eqnarray*}
 t_m \, \, = &  \, \sum\limits_{i=0}^{\lfloor m/2 \rfloor -1} t_i \cdot t_{m-i-1}
& \hbox{if $m$ is even,} \\
t_m \,\, = & \, \binom{t_{(m-1)/2}}{2} +
 \sum\limits_{i=0}^{\lfloor m/2 \rfloor-1} t_i \cdot t_{m-i-1}
 & \hbox{if $m$ is odd}.
 \end{eqnarray*}
 This  holds because each tree $T$ in $\mathcal{T}_m$
 is  constructed uniquely by taking an unordered pair
 consisting of a  tree $T_1$ in $\mathcal{T}_i$
 and a tree $T_2$ in $\mathcal{T}_{m-i-1}$
 and attaching them to a new root. Note that $t_0 = 1$
 corresponds to the case when the new root has
 outdegree one. We call 
 $ \binom{m-1}{i}$ the {\em order} of the
 root. The node is called {\em balanced} if
 $i=(m-1)/2$ and the two subtrees $T_1$ and $T_2$ are isomorphic.
 In this manner, each node of a tree $T \in \mathcal{T}_m$
 has an  order, and it is either balanced or unbalanced.
For instance, all leaves are balanced of order $1$,
all nodes with one child are unbalanced of order $1$, and
nodes with two children have order $\geq 2$.
For a tree $T \in \mathcal{T}_m$ let
${\rm unbal}(T)$ denote the number of unbalanced nodes in the tree $T$, and
let ${\rm order}(T)$ denote the product of the orders of all nodes in~$T$.

\begin{thm} \label{thm:formula}
The count generating function for the cyclohedron test equals
$$ \Gamma_n(t) \quad = \quad n \cdot \sum_{T \in \mathcal{T}_{n-1}}
2^{{\rm unbal}(T)} \cdot t^{{\rm order}(T)}. $$
\end{thm}

\begin{proof}
Every signature $\,\sigma = \{\sigma_1,\ldots,\sigma_{n-1}\} \, $ in $\, \mathcal{C}_n$
maps to an unordered tree $T = T(\sigma)$ in $\mathcal{T}_{n-1}$.
If $n=2$ then $T$ is the tree with one node. For $n \geq 3$ we construct
$T$ iteratively as in Algorithm \ref{cyctestalg}: by induction,
the sets {\rm Left} and {\rm Right} correspond to two subtrees $T_1 $ and $T_2$,
and a new root is attached to form the tree corresponding to
$\,\{\delta_i\} \cup  {\rm Right} \cup {\rm Left} $. The order of the
resulting tree $T(\sigma)$ equals the permutation count
$\mathbf{c}(\sigma)$ computed.
It remains to be shown that the set of all signatures
$\sigma$ which are mapped to the same tree $T \in \mathcal{T}_{n-1}$
has precisely  $\,n \cdot 2^{{\rm unbal}(T)}\,$ elements.
The factor $n$ comes from the fact that the last
element $\delta_n$ can be chosen arbitrarily. So, let us suppose $\delta_n = n$.
Then the indices appearing in $\sigma$ are precisely $1,2,\ldots,n{-}1$.
Let $T_1$ and $T_2$ be the two subtrees of the root
of $T$, and suppose they
have $i$ and $n-2-i$ nodes respectively.
If $i \not= n/2$ then either $\delta_{n-1} = i+1$ and both
$\{1,2,\ldots,i\}$ and $\{n{-}2{-}i,\ldots,n{-}2,n{-}1\}$
are in $\sigma$, or
$\delta_{n-1} = n{-}1{-}i$ and both
$\{1,2,\ldots,n{-}2{-}i\}$ and
$\{n-i,\ldots, n{-}2 , n{-} 1 \}$ are in $\sigma$.
If $i=n/2$ then $\delta_{n-1} = n/2$ and both
$\,\{1,2,\ldots,n/2-1\} \,$ and $\,\{n/2+1,\ldots,n-1\}\,$ are in $\sigma$.
The choices for the remaining elements of $\sigma$ are
constructed inductively by identifying the nodes
of the two subtrees with these two sets.
If the two subtrees are identical (i.e.~the root is balanced)
then there is only one identification to be considered,
otherwise we must consider two cases.
Proceeding in this manner along the tree, we see
that there are $2^{{\rm order}(T)}$ many
choices of signatures $\sigma$ on $\{1,2,\ldots,n-1\}$
which map to $T$.  \end{proof}

We next present a recursive method for computing the
count generating function $\Gamma_n(t)$.
Let $f = \sum_i a_i t^i \,$ and
$g = \sum_j b_j t^j$ be any two generating functions
and $M$ any positive integer. Then we define the {\em $*$-product}
of $f $ and $g$ with respect to $M$ as follows:
\begin{equation}
f  *_M g \quad := \quad \sum_{i,j} a_i \cdot b_j \cdot t^{i\cdot j \cdot M}.
\end{equation}

\begin{cor} \label{cor:recursion}
Let $\Omega_n(t)$ be the polynomial defined recursively by
$$ \Omega_0(t) \,=\, \Omega_1(t) \,=\, t \qquad \hbox{and} \qquad
     \Omega_m (t) \,\,=\,\, \sum_{i=0}^{m-1} \Omega_{i}(t) *_{\binom{m-1}{i}} \Omega_{m-1-i}(t). $$
 Then $\, \Gamma_n(t) \, = \, n \cdot \Omega_{n-1}(t)\,$ is the
 count generating function  for the cyclohedron test.
 \end{cor}

\begin{proof}  This follows from the recursive tree construction
in the proof of Theorem \ref{thm:formula}.
\end{proof}

\begin{example}
Corollary \ref{cor:recursion} easily yields
the full expansion of $\Omega_n(t)$ for small values of $n$.
For $n=17$, the case of interest in
Section 4 (see also Examples \ref{ex:obox} and \ref{ex:obox2}), we find
\begin{eqnarray*}
 \Gamma_{17}(t)  & = \,\,\,\,
 272 t^{108108000} + 544 t^{89689600}+
 272 t^{86486400}+544 t^{80720640} + \cdots
\,\cdots \,  +348160t^8 \\ &\,\,\,\,
+278528 t^7+417792 t^6+278528 t^5+278528 t^4+278528 t^3+139264 t^2+557056 t.
 \end{eqnarray*}
 The number of terms in this polynomial equals
 $\,|\im(\mathbf{c})| \,= \, 2438 $. The $2438$ values
 of the probability distribution function $P_{\mathbf{c}}$
 are plotted on a logarithmic scale in Figure 4.
For larger values of $n$, say $n \geq 30$, it becomes infeasible to compute
the expansion of $\Gamma_n(t)$, but Corollary
\ref{cor:recursion} can still be used to design efficient methods for
sampling from $P_{\mathbf{c}}$. \qed %Referee didn't like this statement.
\end{example}
The distribution function $F_{(r)}(\gamma)$  of the order statistic $\,{\bf C}_{(r)}$ is now computed.
Defining $\,p_i \,= \, P_\mathbf{c}(\gamma_i)$ to be the probability under the null hypothesis that the count is equal to $\gamma_i$,  Remark \ref{rem:genf} tells us that
$$\, (t/n !) \cdot \Gamma'_n(t) \,\,\, = \,\,\,
p_1 t^{\gamma_1} + p_2 t^{\gamma_2} + \cdots + p_{s_n} t^{\gamma_{s_n}},
\qquad \hbox{where} \quad
\gamma_1 < \gamma_2 < \cdots < \gamma_{s_n}. $$
Consider the identity
$$(p_1+ p_2 + \cdots + p_{s_n})^N \quad = \,\,\,
\sum_{i_1+ i_2+ \cdots+i_{s_n} = N} {N \choose i_1 \, i_2 \, \cdots \, i_{s_n}}
\cdot p_1^{i_1} p_2^{i_2} \cdots p_{s_n}^{i_{s_n}} \quad = \quad  1.$$
By definition, $F_{(r)}(\gamma_k)$ is the sub-sum of all terms in
this sum whose indices satisfy
$$\,i_1 + \cdots + i_{k-1} \,\,< \,\, r \,\,\leq \,\,N - i_{k+1} - \cdots -i_{s_n}. $$
For the purpose of computational efficiency we rewrite this sub-sum as follows.
The formula below furnishes us with an efficient method for computing the collective p-values.

\begin{lem} \label{lem:orderstat}
The probability distribution function under the null distribution $Q_0$ of the order statistic ${\bf C}_{(r)}$
is given by
 $$ F_{(r)}(\gamma_k) \,\,\, = \,\,\,\,
 \sum_{i=1}^{N} \sum_{j=\max(0,r-i)}^{\min(r-1, N-i)} \binom{N}{i , j , N {-} i {-} j}
(p_1+\cdots+p_{k-1})^j \cdot p_k^i \cdot
(p_{k+1} + \cdots + p_{s_n})^{N-i-j}
$$
\end{lem}

\begin{proof}
The first sum is over the number of data points that have permutation count $\gamma_k$.
 The second sum is over $j$, the number of data points whose
 permutation count is less than $\gamma_k$.  Then, the multinomial coefficient gives the possible ways to partition
 $\{1,2,\ldots,N\}$ into sets of size $i$, $j$, and $N-i-j$; that is, it accounts for possible rearrangements among the permutation counts  equal to, less than, and greater than $\gamma_k$.
 The probability that such rearrangement occurs is the product
$\, (p_1+\cdots+p_{k-1})^j \cdot p_k^i \cdot
(p_{k+1} + \cdots + p_{s_n})^{N-i-j}$.
\end{proof}

\section*{Appendix: R code for the cyclohedron test}
The {\tt R} source code {\tt topoGraph.R} is available for the
cyclohedron test.  The software can be
downloaded from
\begin{center}
{\tt http://bio.math.berkeley.edu/ranktests/index.html}
\end{center}
Our code requires the free statistical software package {\tt R} \cite{R}.
Here we describe how to perform basic tasks related to the cyclohedron test.  
The data
file must be a CSV
(comma-separated values) file, where the first column consists of
identifying labels (such as gene names), and the first row labels the time points (all other rows are
the corresponding data vectors).  We illustrate the use of the basic functions with the data file  (named `13873.csv')
that we described in Section \ref{Mouse}.  The first column consists of the ProbeSet IDs. 
The source 
code containing the {\tt R} functions is {\tt topoGraph.R}.
 First, we call the source code and load
the data file from an {\tt R} command line (here, we assume that both files are in the current working directory):

\begin{verbatim}
source("topoGraph.R")
dataset<-loaddata("13873.csv")
\end{verbatim}
Next, we calculate the count of each data vector, which is done by the following command:
\begin{verbatim}
counts<-cycleCounts(dataset)
\end{verbatim}
This defines ``counts,'' a vector which lists
the counts $\mathbf{c}$ of the data vectors in the order given by the data file.  To list the genes according to
their count ranking, as shown in Table 1, we call the function {\tt
  rankby} which outputs the labels (here, the ProbeSet IDs) of the genes.  The following command outputs the ten highest ranked ProbeSets.
  
\begin{verbatim}
rankby(row.names(dataset),counts)[1:10]
 [1] "1456017_x_at" "1452041_at"   "1418593_at"   "1417985_at"  
 [5] "1436845_at"   "1436343_at"   "1426267_at"   "1420360_at"  
 [9] "1449643_s_at" "1417399_at"
\end{verbatim}
More extensive documentation is available online.
%
%Other topographical models can also be performed.
%For example, the associahedron test (the graphical model corresponding
%to the path) \cite{semigraphoid} can be performed, by first specifying the graph $G$ and
%then calculating the counts:
%\begin{verbatim}
%G<-buildGraph(17,`path')
%counts2<-preimagesizes(dataset,G)
%\end{verbatim}
%Code for calculating p-values not yet available.
 
 \section*{Acknowledgments}
We are grateful to Mary-Lee Dequ\'{e}ant and Olivier Pourqui\'{e} for
many helpful discussions, and to Mary-Lee for preparing the data for our
use.  We thank Oliver Wienand for helping us implement the cyclohedron test.  The collaboration was facilitated by the  DARPA Fundamental Laws of Biology
Program which supported our research.  Anne Shiu was supported by a Lucent Technologies Bell Labs Graduate Research Fellowship.   We thank an anonymous referee for helpful comments.

% BEPRESS wants the bibliography on the same page.
%\newpage

\end{document}